# Intertwined magnetoresistance and Hall multifunctionality in a non-coplanar magnetic Weyl semimetal $DyB_4$

*Long Chen, Yulin Shen, Songxue Chi, Seunghoon Song, Yang Zhang*, Jian Liu*, and Haidong Zhou**

L. Chen, Y. L. Shen, S. H. Song, Y. Zhang, J. Liu, H. D. Zhou
Department of Physics and Astronomy, University of Tennessee, Knoxville, TN, USA
E-mail: yangzhang@utk.edu; jianliu@utk.edu; hzhou10@utk.edu.

S. X. Chi
Neutron Scattering Division, Oak Ridge National Laboratory, Oak Ridge, Tennessee 37831, USA

Funding: the Air Force Office of Scientific Research under grant no. FA9550-23-1-0502

**Abstract:** Anomalous magneto-transport responses provide complementary probes of orbital motion, momentum-space topology, and real-space spin chirality, yet their integration into a single material remains rare because their underlying requirements often compete. A promising materials-design strategy is to realize a magnetic Weyl semimetal that combines linearly dispersive high-mobility bands with tunable non-coplanar magnetism while limiting spin-dependent scattering. Here we identify $DyB_4$, a frustrated rare-earth tetraboride, as a magnetic Weyl semimetal candidate that embodies this strategy and hosts intertwined magnetoresistance and Hall multifunctionality. Neutron diffraction reveals a sequence of field-tunable magnetic states, including non-coplanar spin configurations and *PT*-symmetry-broken phases. First-principles calculations identify steep linear dispersions and field-induced Weyl points near the Fermi level. Magneto-transport measurements establish a rare fourfold combination of extremely large magnetoresistance, chiral-anomaly-like negative magnetoresistance, large anomalous Hall conductivity arising from cooperative intrinsic Berry curvature and skew scattering, and scalar-spin-chirality-driven topological Hall responses. This multifunctionality arises from the distinct yet weakly coupled roles of itinerant carriers and localized 4*f* moments, which enable high-mobility transport, field-induced Weyl topology, and non-coplanar magnetism. $DyB_4$ therefore provides a 4*f*-electron platform for correlating orbital transport, momentum-space Berry curvature, and real-space spin chirality, suggesting a route toward multifunctional magnetic topological materials.

## 1. Introduction

Magneto-transport, which probes the response of charge transport to an external magnetic field, is a fundamental tool for characterizing electronic materials, evaluating their functional performance, and revealing unconventional electronic states.[1-2] Comprising magnetoresistance (MR) and the Hall effect, it provides key insights into carrier dynamics, scattering mechanisms, band structure, Fermi-surface properties, electronic correlations, and the coupling between conduction electrons and magnetic order.[2-5] Recent studies have uncovered a variety of anomalous magneto-transport responses, including large magnetoresistance (LMR),[6] whose extreme form is known as extremely large magnetoresistance (XMR),[7-9] negative magnetoresistance (NMR),[10] the anomalous Hall effect (AHE),[4] and the topological Hall effect (THE).[11-12] Originating from distinct orbital, momentum-space, and real-space mechanisms, these responses offer complementary functionalities, ranging from high-sensitivity magnetic sensing and anisotropic transport to electrical readout of magnetic order and topological spin textures.[1, 13-17]

Although one or a few anomalous magneto-transport responses have been observed in various systems,[18-40] their simultaneous realization in a single material remains quite rare (**Figure 1a**). This scarcity reflects the distinct and often competing physical requirements underlying each magneto-transport signature (Figure 1b). Intrinsic AHE, characterized by anomalous Hall conductivity (AHC, $\sigma^{A}_{xy}$), and chiral-anomaly-induced NMR are closely associated with Weyl-related topology through Berry curvature[41-42] and chiral charge dynamics,[43-45] whereas XMR generally requires high-mobility electronic states[23] or a nearly compensated electron-hole scenario.[7-8] In contrast, THE, characterized by topological Hall resistivity ($\rho^{T}_{xy}$) originates from non-coplanar magnetic textures,[46-47] which are typically fragile and difficult to tune without introducing additional scattering.[11, 48-49] In magnetic systems, spin-dependent scattering, such as the field-induced suppression of spin-disorder or magnetic-fluctuation scattering, can also give rise to extrinsic NMR,[50-52] thereby obscuring both XMR and intrinsic chiral-anomaly-induced NMR. Therefore, a single material hosting all four effects would not only provide a rare platform for disentangling orbital transport, momentum-space topology, and real-space spin textures, but also establish a multifunctional quantum-material platform by integrating these complementary anomalous magneto-transport responses.

From a materials-design perspective, realizing such a system requires a magnetic Weyl material that combines high-mobility or electron-hole-compensated electronic states with tunable non-coplanar spin textures, while minimizing spin-related scattering in high-quality crystalline samples. Unlike accidental and difficult-to-control electron-hole compensation, high mobility can be more directly promoted by steep, linearly dispersive electronic bands.[53-54] Rare-earth-based magnetic Weyl semimetals offer a promising platform in this regard: their relatively extended 4*d* orbitals,[55] often hybridized with ligand *p* orbitals, may promote broader band dispersion and could support linear band crossings over a wider energy range, particularly compared with 3*d* transition-metal systems with analogous crystal structures. Meanwhile, localized 4*f* moments provide strong magnetic anisotropy[56] and possible multipolar interactions,[57-58] which may stabilize field-tunable non-collinear or non-coplanar spin textures.[59-62] The localized nature of 4*f* moments also limits their coupling with conduction carriers, thereby reducing the extent to which magnetic order perturbs the itinerant bands.[56, 63-64] In this work, we report the coexistence of all four anomalous magneto-transport behaviors in $DyB_4$, a newly identified magnetic Weyl semimetal candidate featuring a high-mobility small hole pocket and tunable non-coplanar spin textures. These results establish $DyB_4$ as a compelling platform for correlating orbital transport, momentum-space topology, and real-space spin textures, while also highlighting its potential for electronic and spintronic multifunctionalities.

## 2. Multiple magnetic states and corresponding AHE and spin chirality

$DyB_4$ is a frustrated rare-earth tetraboride crystallizing in a tetragonal nonsymmorphic structure (space group $P4/mbm$), where $Dy^{3+}$ ions form a Shastry-Sutherland lattice (SSL) (**Figure 2**a) and the interplay of frustration, Ruderman-Kittel-Kasuya-Yosida (RKKY) interactions, and quadrupolar-lattice coupling drives successive magnetic and quadrupolar orderings with strong field tunability.[65-68] According to detailed magnetic susceptibility and magnetization measurements (see Supplementary Note 1, Figure S1-S3), the corresponding magnetic field-temperature ($H$-$T$) phase diagrams have been established. As shown in Figure 2b-c, paramagnetic (PM) $DyB_4$ undergoes an antiferromagnetic ($T_N$ = 21 K) and a ferroquadrupolar ordering (FQO, $T_Q$ = 13 K) upon cooling under zero magnetic field. At the lowest temperature ($T$ = 2 K), the out-of-plane magnetic field drives $DyB_4$ through a magnetization plateau phase and eventually into a polarized ferromagnetic state. The plateau phase further evolves into a separate region at relatively higher temperatures, which is consistent with previous reports.[67] In contrast, the in-plane magnetic field only suppresses the FQO, whereas the intermediate antiferromagnetic state remains almost unaffected up to 14 T (Figure S3).

Previous neutron and resonant X-ray diffraction studies have established the zero-field collinear antiferromagnetic (CAFM) state and the associated non-coplanar antiferromagnetic (NCAFM) ground state that emerges concomitantly with FQO.[61-62, 65] To resolve the field-driven evolution of the magnetic structures in $DyB_4$, we performed low-temperature neutron diffraction measurements under a vertical magnetic field with the single-crystalline sample aligned in the ($HK$0) scattering plane (Figure S4). As shown in Figure 2d, the field dependence of the integrated intensities of several reflections closely tracks the magnetization process and displays a pronounced plateau-like stage over the field range corresponding to the magnetization plateau phase. At high magnetic fields, the integrated intensities of the (100) and (300) magnetic reflections are strongly suppressed and approach zero, indicating the formation of a fully polarized collinear ferromagnetic (CFM) state with no detectable in-plane moment. This assignment is consistent with the neutron magnetic scattering selection rule, whereby only magnetic moment components perpendicular to the scattering vector $Q$ contribute to the diffraction intensity. Notably, the (100) intensity decreases continuously from a finite value in the plateau phase to nearly zero in the fully polarized CFM state over a sizable field window of $\Delta H \sim 1$ T. This continuous evolution provides clear evidence for an intermediate non-coplanar ferromagnetic (NCFM) configuration that separates the plateau phase from the fully polarized CFM state. Moreover, the finite (100) magnetic reflection within the low-temperature plateau phase demonstrates that this phase still retains the non-coplanar spin component (Figure 2e); therefore, we identify it as a non-coplanar plateau (NCP) phase. Since the fully polarized CFM state is incompatible with quadrupolar order, the phase boundary between the two plateau regimes marks the suppression of quadrupolar order. Consequently, the remaining plateau region at higher temperature is assigned to a collinear plateau (CP) phase. These field-induced magnetic configurations and phase assignments are further supported by the neutron diffraction refinements (see Supplementary Note 2, Table S1, Figures S5-S10).

Figure 2f summarizes the magnetic structure and the corresponding magnetic space group (MSG), AHE tensor, and the scalar spin chirality of each state (see Supplementary Note 3, Table S2). It should be noted that *PT* symmetry is broken in both the plateau (NCP and CP) and ferromagnetic (NCFM and CFM) states, yielding a finite Berry curvature and, consequently, a possible AHE. Motivated by the non-coplanar spin configurations, we further evaluated the scalar spin chirality, $\chi_{ijk} \equiv S_i \cdot (S_j \times S_k)$, in each magnetic state. The nonzero scalar spin chirality appears in the NCP and NCFM states, implying that these states may give rise to a THE (Table S2).

**3. Steep linear dispersion and magnetic-field-induced Weyl semimetallic state**

To assess whether $DyB_4$ can host a time-reversal-symmetry-breaking magnetic Weyl state and an associated AHE, the electronic structure of different magnetic states was investigated using first-principles calculations (**Figure 3**a, Figure S11). In the PM state, the bands near $E_F$ are dominated by hybridized Dy-4*d* and B-2*p* orbitals and form a large electron pocket centered at the Γ point. Along the Γ-M direction, a Dirac-cone-like feature exists slightly below $E_F$, resulting in a small electron pocket and a relatively large hole pocket. Together with the sharp bands crossing $E_F$, these features suggest a robust metallic nature for $DyB_4$. Moreover, the smaller electron pocket shrinks substantially and eventually disappears in the NCAFM state (Figure 3b). In all the PM, CAFM, and NCAFM states, the preservation of *PT* symmetry precludes the existence of Weyl points.

However, the breaking of *PT* symmetry in the field-induced plateau (NCP and CP) and ferromagnetic (NCFM and CFM) states may induce Weyl points. For simplicity, we focus on the electronic structure of the NCFM and CFM states, both exhibiting pronounced band splitting and energy shifts compared with the NCAFM state. The loss of time-reversal symmetry leads to the appearance of several Weyl points (Figures S11-S12), most of which are located along the X-M and A-M directions (Tables S3-S4). In particular, the Weyl points in the CFM state lie close to $E_F$, making their contribution to the low-energy electronic structure more prominent. As shown in Figure 3c, the red and black dots denote the Weyl points with positive and negative chiralities, respectively, which are identified from the Wilson loop calculations (Figure 3d). These bulk Weyl points are distinct from the previously proposed magnetic wallpaper fermions in the surface states of $DyB_4$ associated with a noncollinear AFM state in a virtual $\Gamma_4$ phase (MSG: *P*4'/*m'b'm*).[69] These Weyl points naturally generate nonzero Berry curvature and can therefore give rise to an intrinsic AHE. As shown by the chemical-potential-dependent AHC of $DyB_4$ in the CFM state (Figure 3e), $\sigma_{xy}$ exhibits a highly nonmonotonic dependence on energy. The pronounced peaks occur near the energies of the Weyl points, consistent with the enhanced Berry curvature expected around Weyl nodes. Theoretically, the AHC is estimated to range from 403.33 to 613.79 $\Omega^{-1}$ $cm^{-1}$ near $E_F$ in the CFM state (Table S5), comparable in magnitude to that of FM $PrB_4$ (see Supplementary Note 4 for details of the calculations).[70] The presence of steep low-energy bands and field-induced Weyl nodes provides a favorable electronic structure with unconventional magneto-transport properties in $DyB_4$, motivating the exploration of large orbital MR, AHE, and chiral-anomaly-like NMR.

## 4. Extremely large transverse MR and chiral-anomaly-like NMR

The resistivity of $DyB_4$ decreases nearly linearly upon cooling, with a pronounced drop below $T_N$ and a clear kink around $T_Q$ (**Figure 4**a and Figure S13). Upon application of an out-of-plane magnetic field ($H$ // $c$), both the NCAFM and CAFM states are suppressed, accompanied by a significant increase in the low-temperature resistivity, which eventually exhibits metal-insulator-like behavior. When the magnetic field is applied in-plane and still perpendicular to the current ($H$ // $a$, $I$ // $b$), similar metal-insulator-like behavior is observed under high magnetic fields, although the CAFM state is barely suppressed (Figure 4b). In contrast, when the in-plane magnetic field is parallel to the current ($H$ // $b$), the low-temperature resistivity always decreases upon cooling, and no metal-insulator-like behavior appears (Figure 4c). The corresponding transport phase diagrams are consistent with those obtained from magnetic measurements, except for the metal-insulator-like behavior (see Supplementary Note 5, Figure S14).

When the magnetic field is perpendicular to the current, the observed metal-insulator-like behavior corresponds to LMR at low temperatures (Figure 4d,e), reaching maximum values of 17258% at 3 K under 14 T for $H$ // $c$ and 21881% at 2 K under 14 T for $H$ // $a$, respectively. Such an LMR exceeds $10^3$% and remains nonsaturating under high magnetic fields (up to 14 T), fulfilling the criterion for XMR.[9] With increasing temperature, the XMR for both $H$ // $c$ and $H$ // $a$ decreases significantly and monotonically, reaching only a small positive value of approximately 5% at 300 K (Figure 4f). The always-positive MR indicates that the orbital motion of charge carriers under the Lorentz force is dominant,[2] whereas the spin-dependent scattering usually observed in magnetic materials[50-52] is not the primary mechanism for explaining the MR of $DyB_4$. Notably, when the in-plane magnetic field is parallel to the current ($H$ // $b$), the XMR is strongly reduced (down to 150% at 3 K under 14 T), accompanied by pronounced NMR features (Figure 4d). The NMR features emerge under high magnetic fields below $T_N$ and persist over the full magnetic field range slightly above $T_N$ (Figure 4e), corresponding to $PT$-symmetry breaking driven by the suppression of the FQO-associated NCAFM state and the alignment of PM moments under in-plane magnetic fields, respectively. The fact that the NMR features are only observed in the $H$ // $I$ configuration (Figure 4c, Figure S15) under broken $PT$ symmetry suggests that they may originate from the chiral-anomaly effect associated with Weyl points.[43-44] To further verify this behavior, the current direction was changed to the $c$-axis, and clear NMR features were observed only under $H$ // $c$, which also emerge under high magnetic fields below $T_N$ and persist over the full magnetic-field range slightly above $T_N$ (Figure 4g, Figure S16). This strongly supports the interpretation that the NMR is associated with magnetic-field-induced Weyl points through the chiral anomaly effect. The current-jetting effects can be excluded based on the angular-dependent MR measurements (Figure S15) and the observation of NMR in both spine and edge contact geometries (Figure S17).[71-73] These results indicate that the coexistence of XMR and chiral-anomaly-like NMR is an intrinsic magneto-transport response of $DyB_4$.

To the best of our knowledge, the coexistence of XMR and chiral-anomaly-like NMR has been reported only in a limited number of systems (Figure 4h, Table S6), most of which are nonmagnetic topological semimetals with broken inversion symmetry but preserved time-reversal symmetry, such as $\alpha$-As,[18] TaAs,[20] $TaAs_2$,[21] and $Cd_3As_2$.[22-23] In magnetic topological systems with broken time-reversal symmetry, unavoidable spin-related scattering generally competes with positive orbital MR, thereby making the emergence of XMR less favorable. These magnetic topological systems with chiral-anomaly-induced NMR usually show relatively modest positive MR, as exemplified by $Co_3Sn_2S_2$ (220% at 2 K under 9 T),[9, 74] GdPtBi (285% at 2 K under 14 T).[9, 75] The recent observation of an XMR of approximately 2500% at 2 K under 14 T in ultra-high-quality $Co_3Sn_2S_2$ single crystals further highlights that suppressing carrier scattering is a prerequisite for realizing XMR in magnetic topological materials.[36] This can be achieved either extrinsically by improving crystal quality or intrinsically by reducing the coupling between itinerant carriers and magnetic moments. A related route is found in compensated rare-earth monopnictides, such as NdSb and CeSb,[24-25] where localized 4$f$ moments are introduced while preserving the high-mobility semimetallic bands, as well as in the rare-earth tetraboride $GdB_4$, where coexisting XMR and chiral-anomaly-like NMR have been reported.[76] In this context, $DyB_4$ stands out as another distinct platform where the hybridized Dy-4$d$/B-2$p$ conduction electrons and the highly localized Dy-4$f$ moments are expected to be weakly coupled. Combined with field-induced Weyl points, this provides a favorable condition for the coexistence of XMR and chiral-anomaly-like NMR, making $DyB_4$ a rare magnetic Weyl system for exploring the microscopic mechanisms of exotic MR.

## 5. Hall resistivity and origin of the XMR

In order to explore the mechanism of XMR and possible AHE, the field-dependent Hall resistivity ($\rho_{xy}$) at various temperatures was measured (**Figure 5**a). All the Hall resistivity curves show a negative slope, suggesting that electrons are the dominant charge carriers. Obviously, the overall behavior of $\rho_{xy}$ cannot be simply described using the semi-classical single-band model, and the AHE term should be considered in $PT$-symmetry-broken phases at low temperature. Given the nonlinear field dependence of $\rho_{xy}$, a multi-carrier model[76] was employed to analyze the Hall resistivity. The room-temperature Hall resistivity can be fully reproduced with two electron carriers ($e_1$, $e_2$), whereas a hole-type carrier ($h_1$) must be taken into consideration below 30 K to accurately describe the high-field feature (Figure 5b, Figures S18-S19; see Supplementary Note 6 for a discussion of the necessity and validity of the multi-carrier model). Deviations from this model begin to emerge below $T_N$ when the magnetic field exceeds the upper field boundary of the CAFM phase, where a magnetization-related AHE term starts to appear. Focusing on the low-field Hall resistivity in the CAFM and NCAFM states, we noticed that the nonlinear feature under the positive magnetic field gradually changes from the upturn ("∪") shape to the downturn ("∩") shape when crossing $T_Q$ (Figure 5c). At 7 K, even a simple single band is sufficient to describe the low-field Hall resistivity (dashed line). Despite this evolution, the overall negative slope remains nearly unchanged, indicating that the change after entering the NCAFM state is mainly associated with minority carriers rather than the dominant majority carriers.

Figure 5d-e shows the temperature dependence of the carrier density and mobility for each type of charge carrier. At room temperature, a large electron pocket exists as the majority carrier ($n_{e1} \sim 5 \times 10^{21}$ cm$^{-3}$), accompanied by a small electron pocket ($n_{e2} \sim 3 \times 10^{19}$ cm$^{-3}$). Upon cooling, the densities of both electron carriers decrease, while a small hole pocket ($n_{h1} \sim 1 \times 10^{19}$ cm$^{-3}$) gradually emerges and its density increases. When crossing $T_Q$, the decrease (increase) in the small electron (hole) density accelerates, and both the small electron and hole

pockets vanish simultaneously and are eventually replaced by another small hole pocket ($n_{h2}$ ~ $6 \times 10^{18}$ cm$^{-3}$) (Table S7). Such an evolution may correspond to a Lifshitz transition[77] in light of the band structure of $DyB_4$ (Figure 5f, upper panel), potentially induced by quadrupolar-lattice coupling that modifies both the magnetic and crystal structures.[62] Such a scenario is qualitatively supported by our calculation, in which the PM state exhibits three pockets, and the evolution during the cooling process is consistent with the Fermi-surface reconstruction tendency (Figure 2b).

The obtained carrier densities and mobilities at 2 K are $n_{e1}$ = 5.86(4) × $10^{21}$ cm$^{-3}$, $\mu_{e1}$ = 328(5) cm$^2$ V$^{-1}$ s$^{-1}$, $n_{h2}$ = 6.16(94) × $10^{18}$ cm$^{-3}$, $\mu_{h2}$ = 6075(303) cm$^2$ V$^{-1}$ s$^{-1}$ (Table S7). The hole carrier density is about three orders of magnitude lower than that of the dominant electrons, indicating a substantial deviation from the electron-hole compensation mechanism typically used to account for XMR.[7-8, 76, 78] Nevertheless, the combination of low carrier density and high mobility of the hole pocket suggests an alternative mechanism underlying the observed XMR, especially considering the steep linear dispersion near $E_F$ (Figure 5f, bottom panel). When the magnetic field is large and perpendicular to the current, the Lorentz force significantly bends the carrier trajectories, leading to a strong magnetoresistive effect. In this regime, the longitudinal conductivity can be described by $\sigma_{xx} = \sigma_0/(1+(\omega_c\tau)^2) \approx \sigma_0/(\omega_c\tau)^2$, where $\sigma_0$ is the zero-field conductivity, $\omega_c$ is the cyclotron frequency, and $\tau$ is the carrier relaxation time proportional to the carrier mobility ($\tau = \mu m/e$)[2]. As in our observation, the carrier mobility of the hole pocket is quite high, and once the cyclotron frequency is large enough, an XMR is anticipated. In a linear band dispersion with $E(k) = \hbar v_F \cdot |k|$, the cyclotron frequency is derived to be $\omega_c = eB/m_c = 2\pi eB/\hbar^2 \cdot (\partial A/\partial E)^{-1} = eBv_F/\hbar \cdot k_F^{-1}$, which naturally results in XMR when the carrier pocket is extremely small ($k_F \sim 0$). Thus, unlike the electron-hole compensation mechanism, the XMR observed in $DyB_4$ is more likely to be attributed to steep linear dispersion and the emergence of the high-mobility small hole pocket at low temperatures.

**6. Anomalous Hall resistivity and topological Hall resistivity**

Considering the finite magnetization and nonzero scalar spin chirality under high out-of-plane magnetic fields, the Hall resistivity below $T_N$ can be expressed as $\rho_{xy} = \rho^{OHE}_{xy} + \rho^{AHE}_{xy} + \rho^{T}_{xy}$, where $\rho^{OHE}_{xy}$, $\rho^{AHE}_{xy}$, and $\rho^{T}_{xy}$ denote the ordinary, anomalous, and topological Hall resistivity, respectively. **Figure 6**a shows the decomposition of each term at 2 K. After subtracting the ordinary Hall contribution, the combined anomalous and topological Hall term ($\rho^A_{xy}$) can be obtained (Figure 6b), from which the THE contribution ($\rho^T_{xy}$) is further extracted by subtracting the AHE term ($\rho^{AHE}_{xy} = R_sM$)[4] that is linearly coupled with the magnetization (Figure 6c). In this analysis, the low-temperature ordinary Hall resistivity is described by the multi-carrier model in the low-field region, while an additional field-independent AHE term is included to account for the Hall resistivity in the fully polarized CFM state under high magnetic fields (Figure S20, Supplementary Note 7).

To further evaluate the anomalous Hall response, we first calculate the total AHC from the combined anomalous and topological Hall contribution using $\sigma^A_{xy} = -\rho^A_{xy}/(\rho^2_{xx} + (\rho^A_{xy})^2)$, where $\rho_{xx}$ represents the field-dependent longitudinal resistivity. As the magnetic field increases, the calculated total AHC shows a sudden increase around the plateau/intermediate phase region, followed by a gradual decrease and a tendency toward saturation upon entering the CFM state (Figure S21a). We therefore focus on the total AHC at 14 T, representing the high-field CFM state, as well as the maximum total AHC that emerges near the intermediate phase. At low temperatures, the total AHC at 14 T is already much larger than the value estimated from calculation (~600 $\Omega^{-1}$ $cm^{-1}$) and exceeds the characteristic quantization scale $e^2/ha$ ($10^2$~$10^3$ $\Omega^{-1}$ $cm^{-1}$, $a$ is the lattice constant),[79] reaching 1478 $\Omega^{-1}$ $cm^{-1}$ at 2 K. More remarkably, the maximum total AHC reaches an extremely large value of 14090 $\Omega^{-1}$ $cm^{-1}$ at 2 K. With increasing temperature, both the total AHC at 14 T and the maximum total AHC first show a slight enhancement and then gradually decrease to zero near $T_N$. The total AHC at 14 T reaches a maximum value of 3134 $\Omega^{-1}$ $cm^{-1}$ at 10 K, while the maximum total AHC reaches 16135 $\Omega^{-1}$ $cm^{-1}$ at 3 K (Figure S21b). Such a large value of AHC is among the largest ones observed in the reported AHE systems, comparable in magnitude to those of MnGe thin film[80] (~40000 $\Omega^{-1}$ $cm^{-1}$) and recently discovered diamagnetic $La_3MgBi_5$[81] (42356 $\Omega^{-1}$ $cm^{-1}$). Figure 6d shows the logarithmic plot of $\sigma^A_{xy}$ *vs* $\sigma_{xx}$, together with several representative AHE materials.[35, 80-87] The AHC at 14 T follows the scaling relation with an exponent approaching 1 ($\alpha$ = 0.94(1)), suggesting that the AHC in the CFM state is dominated by skew scattering.[79, 86] Considering the magnetic-field-induced Weyl points in the NCFM and CFM states, the large AHC observed in $DyB_4$ likely arises from a rare cooperative mechanism between skew scattering and field-induced momentum-space Berry curvature. Considering the maximum total AHC, $DyB_4$ exhibits a giant AHE despite lying in the moderately conductive regime, occupying an unusual region of the scaling plot together with $La_3MgBi_5$,[81] MnGe thin film,[80] $KV_3Sb_5$,[83] and $CsV_3Sb_5$.[82] The cooperation between skew scattering and intrinsic momentum-space Berry curvature has also been suggested in $La_3MgBi_5$[81] and $CsV_3Sb_5$,[88] and the appearance of $DyB_4$ in this unusual region further indicates that such cooperation may represent an unconventional AHC regime and provide a possible route for searching for large-AHE materials.

To separate the contribution from the anomalous Hall term, we also calculated the pure AHE-derived AHC using $\sigma^{AHE}_{xy} = -\rho^{AHE}_{xy}/(\rho^{2}_{xx} + (\rho^{AHE}_{xy})^2)$, which shows field and temperature dependencies similar to those of the total AHC (Figure S22). However, a clear difference appears in the maximum AHC, where the pure AHE-derived maximum AHC is much larger than the maximum total AHC, reaching 20332 $\Omega^{-1}$ $cm^{-1}$ at 3 K. This difference indicates that the topological Hall contribution does not enhance the anomalous Hall response in $DyB_4$. Instead, it partially compensates for the AHE contribution, most likely because the THE term has an opposite sign to the AHE term in the relevant field range. In the present scattering-dominated regime, asymmetric carrier scattering associated with chiral spin textures may contribute to the THE, thereby offsetting the skew-scattering-dominated AHE. Such compensation may suggest a complex interplay between the field-induced momentum-space Berry curvature contribution and the spin-chirality-induced real-space Berry curvature contribution in the anomalous Hall response, which deserves further theoretical and experimental investigation to clarify their respective roles.

Besides, the topological Hall resistivity is extracted to be approximately 0.3 $\mu\Omega$ cm at 2 K, larger than or comparable to that of several early-discovered skyrmion-hosting systems (Figure 6e, Table S8), such as MnSi (< 0.04 $\mu\Omega$ cm[11, 89]), FeGe (-0.136 $\mu\Omega$ cm[90]), MnGe (-0.16 $\mu\Omega$ cm[91]), and $Fe_3Sn_2$ (-0.875 $\mu\Omega$ cm[33]), et al. However, it is smaller than those reported in non-coplanar spin systems, including $LaMn_2Ge_2$ (1.5/4.5 $\mu\Omega$ cm[92-93]), $Gd_2PdSi_3$ (2.6 $\mu\Omega$ cm[59]), $Fe_3GeTe_2$ (2.04 $\mu\Omega$ cm[94]), $MnBi_4Te_7$ (7 $\mu\Omega$ cm[95]), and Heusler alloy $Ni_2Mn_{1.4}In_{0.6}$ (6 $\mu\Omega$ cm[96]). Phenomenologically, the magnitude of the THE is governed by both the scalar spin chirality and the coupling strength between conduction electrons and localized magnetic moments. In skyrmion-hosting systems such as MnSi and FeGe, the long-wavelength ($\lambda_{sk}$ = 18 - 70 nm)[11, 97] spin texture leads to a relatively small spin-chirality density and hence a small emergent magnetic field, which scales approximately as $B_{\mathrm{em}} = -h/(2/\sqrt{3}e\lambda_{sk}^2)$,[59] where $\lambda_{sk}$ is the characteristic length scale of the spin texture. This explains why long-period skyrmion phases usually exhibit relatively small THE signals.[91] In contrast, static non-coplanar spin systems often host spin textures confined within only a few unit cells ($\lambda_{sk}$ ~ 1 nm), producing a much larger spin-chirality density and therefore a larger THE. For $DyB_4$, the relevant magnetic modulation is characterized by $\boldsymbol{q}$ = (0, 0, 0),[62] corresponding to a characteristic length scale of approximately $\lambda_{sk} \sim \boldsymbol{a}$ = 0.7 nm. This length scale is much smaller than those of $LaMn_2Ge_2$ ($\boldsymbol{q}$ = (0, 0, 0.295), $\lambda_{sk}$ ~ 3.6 nm)[93] and $Gd_2PdSi_3$ ($\boldsymbol{q}$ = (0.141, 0, 0), $\lambda_{sk}$ ~ 2.5 nm).[59] Therefore, $DyB_4$ would be expected to generate a much larger emergent magnetic field based solely on the spin-texture length scale. However, the experimentally observed THE is not as large as expected. This discrepancy may be attributed to the relatively weak coupling between the highly localized Dy-4*f* moments and the hybridized Dy-4*d*/B-2*p* conduction electrons, in contrast to the stronger *d*-*d* or *p*-*d* coupling commonly found in transition-metal-containing non-coplanar spin systems.

**7. Discussion and Conclusions**

The present results establish $DyB_4$ as an unusual magnetic Weyl semimetal candidate in which multiple anomalous magneto-transport responses coexist and can be tuned among distinct magnetic states. Although the coexistence of two, and occasionally three, anomalous magneto-transport responses has been reported in several materials, $DyB_4$ achieves a rare fourfold combination by integrating XMR, chiral-anomaly-like NMR, large AHC, and a moderate topological Hall resistivity within a single material. Such a combination not only offers potential opportunities for multifunctional electronic and spintronic applications but also provides an important platform for understanding competing and cooperative mechanisms underlying magneto-transport properties.

A key ingredient in these intertwined anomalous magneto-transport responses in $DyB_4$ is the distinct yet weakly coupled roles played by the itinerant Dy-4*d*/B-2*p* bands and the highly localized Dy-4*f* magnetic moments. The itinerant Dy-4*d*/B-2*p* bands form steep linear dispersion and generate a small high-mobility hole pocket at low temperatures, providing a favorable electronic environment for large orbital MR. Rather than conventional electron-hole compensation, the combination of a small high-mobility hole pocket and linear band dispersion in $DyB_4$ offers a more natural explanation for the nonsaturating XMR. Meanwhile, the localized Dy-4*f* moments stabilize multiple field-tunable magnetic states, including both *PT*-symmetry-breaking phases with a nonzero AHE tensor and non-coplanar configurations with finite scalar spin chirality. The coupling between the itinerant Dy-4*d*/B-2*p* bands and the highly localized Dy-4*f* magnetic moments allows the magnetic order to reconstruct the electronic topology and generate field-induced Weyl points near $E_F$, providing a natural origin for the observed chiral-anomaly-like NMR. At the same time, its weak nature suppresses strong spin-dependent scattering, creating relatively favorable conditions for observing nonsaturating XMR and intrinsic chiral-anomaly-like NMR. Notably, the coexistence of XMR and chiral-anomaly-like NMR in $DyB_4$ produces a strong contrast between transverse and longitudinal MR, suggesting the possibility of exceptionally large angular-dependent MR. This result provides a novel design principle for realizing large angular-dependent MR by combining high-mobility bands for enhanced transverse XMR with field-induced Weyl points for longitudinal chiral-anomaly-like NMR.

The anomalous Hall response further reveals the cooperative roles of momentum-space topology, real-space topology, and scattering mechanisms. Experimentally, $DyB_4$ exhibits a large AHC that substantially exceeds the intrinsic value estimated from Berry-curvature calculations. The scaling analysis indicates a dominant skew-scattering contribution in the high-field CFM state, suggesting that the large AHE likely reflects a cooperative contribution from intrinsic momentum-space Berry curvature and extrinsic skew scattering, rather than a purely intrinsic Berry curvature response. This places $DyB_4$ in an unconventional anomalous Hall regime that goes beyond a simple intrinsic or extrinsic classification. In addition to the AHE, the non-coplanar magnetic phases of $DyB_4$ generate nonzero scalar spin chirality and induce a moderate topological Hall resistivity, consistent with weak coupling between the localized 4*f* moments and conduction electrons. The partial compensation between opposite AHE and THE contributions further indicates a complex interplay between momentum-space Berry curvature and real-space spin chirality, which deserves further microscopic investigations through angle-resolved photoemission spectroscopy, quantum-oscillation measurements, and field-dependent microscopic magnetic-structure studies. Nevertheless, owing to its moderate magnitude, THE only weakly reduces the total AHC, allowing $DyB_4$ to retain a remarkably large anomalous Hall response. This observation conversely implies that, in systems where the AHE and THE contributions have the same sign, enhancing the coupling between chiral spin textures and itinerant carriers could further amplify the total Hall response through a THE-assisted mechanism.

In conclusion, combining neutron diffraction, first-principles calculations, and detailed magneto-transport measurements, we have demonstrated that $DyB_4$ hosts intertwined magnetoresistance and Hall multifunctionality rooted in the interplay among high-mobility carriers, field-induced Weyl topology, and non-coplanar magnetism. These findings position $DyB_4$ as a model 4*f*-electron platform for disentangling orbital MR, momentum-space Berry curvature, and real-space spin chirality. More broadly, they suggest that rare-earth magnetic Weyl materials, where localized moments and itinerant carriers are weakly but functionally coupled, offer a promising design principle for realizing multifunctional topological transport responses for future electronic and spintronic applications.

### 8. Experimental Section

*Single Crystal Growth and Structure Characterization.* $DyB_4$ single crystals were grown by a high-temperature solution method using Al as flux. The isotope $B^{11}$ was used in the growing process to reduce potential neutron absorption. Square-shaped, shiny-silver single crystals with sizes up to 3 mm × 3 mm × 1 mm can be obtained. X-ray diffraction (XRD) data on a single crystal at room temperature were obtained using a PANalytical Empyrean diffractometer (Cu $K_\alpha$ radiation, $\lambda$ = 1.54178 Å) operated at 45 kV voltage and 40 mA current, with a graphite monochromator in a reflection mode ($2\theta$ = 5°–100°, step size = 0.026°). Indexing and Rietveld refinement were performed using the DICVOL91 and FULLPROF programs.[98] The crystal orientation was determined using the Laue backscattering method.

*Physical Property Measurement.* Temperature-dependent magnetic susceptibility and the field-dependent magnetization curves were measured using a vibrating sample magnetometer system (Quantum Design, Dynacool-14 T). Then, the samples (S#1-S#3) were polished into a rectangular shape with a thickness of about 0.15 mm. Resistance, magnetoresistance, and Hall resistance were measured using the standard four-probe configuration, with the applied current (about 3 mA) along the *a*- (S#1-S#2) and *c*-axes (S#3). Heat capacity measurements were carried out at temperatures ranging from 2 K to 200 K under high vacuum (0.01 $\mu$bar).

*Neutron Single-Crystal Diffraction.* Elastic neutron scattering was conducted on a 10 mg single crystal using the TAX (HB-3) instrument at the High Flux Isotope Reactor of Oak Ridge National Laboratory. The large crystal was polished to a size of ~ 2 × 2 × 0.05 $mm^3$ to allow roughly 10% neutron transmission (wavelength ~ 1 Å). The sample was mounted on an aluminum holder and aligned in the (*HHL*) ($H$ = 0 T) and (*HK0*) ($H$ // $c$) scattering planes within a cryostat with a base temperature of 1.5 K. A neutron wavelength of 1.486 Å was selected using a Ge monochromator. To determine the zero-field magnetic structure, several magnetic Bragg peaks were measured at 25 K, 15 K, and 2 K. Field-dependent measurements ($0 \leq H \leq 8$ T with 0.25 T step) were then carried out by collecting the intensities (*k*-scan) of the (100), (110), (200), and (300) reflections under an out-of-plane magnetic field at 2 K and 15

K. Data reduction was performed using DAVE[99] and code developed based on the TasVisAn and InsPy packages.[100] Symmetry analysis and refinements of the neutron diffraction data were carried out using the SARAh and FullProf suite,[101-102] and the refined structures were visualized with VESTA.[103] The integrated intensities are extracted after removing appropriate resolution-based Lorentz factors, and the scattering data are analyzed by subtracting the 25 K data (above $T_N$) to obtain pure magnetic signals by eliminating lattice contributions.

*First-Principles Calculations.* First-principles calculations were performed using the Vienna *ab* initio Simulation Package (VASP) with the projector-augmented-wave (PAW) method.[104-105] The exchange-correlation interaction was treated within the Perdew-Burke-Ernzerhof (PBE) generalized-gradient approximation (GGA).[106] The on-site Coulomb interaction for Dy-4*f* states was included using the Dudarev GGA + $U$ approach, with $U$ = 9.8 eV and $J$ = 0.7 eV, corresponding to an effective interaction of $U_{eff} = U - J = 9.1$ eV. A small adjustment of the electron count was used to align the calculated Fermi level with the experimentally relevant carrier regime. The 6 × 6 × 12 Γ centered K-point meshes were employed for Brillouin zone (BZ) sampling of the unit cell. For the PM and CAFM states, static calculations were performed using the tetragonal crystal structure with space group $P4/mbm$ (No. 127), $a = b = 7.1082$ Å, $c = 4.0196$ Å, and $\alpha = \beta = \gamma = 90°$.[62] For the NCAFM, NCFM, and fully polarized CFM states of $DyB_4$, the experimental monoclinic lattice parameters with space group $P12_1/a1$ (No. 14), $a$ = 7.1078 Å, $b$ = 7.1086 Å, $c$ = 4.0196 Å, and $\beta$ = 89.82°, were adopted.[62] Spin-orbit coupling (SOC) was included in all calculations. The lattice parameters were fixed to the corresponding experimental structures. Because the tetragonal-to-monoclinic distortion is very small, all band structures were projected along the selected high-symmetry paths in the BZ defined using the tetragonal setting. The maximally localized Wannier functions for 4*d* and 4*f* orbitals of Dy, 2*p* orbitals of B generated by VASP2WANNIER90 and WANNIER90 packages[107] can well reproduce the first-principles electronic band structures. The intrinsic AHC calculation is performed as:

$$\sigma_{yx}^{z}(\mu) = e^2\hbar\left(\frac{1}{2\pi}\right)^3 \int_{\mathrm{BZ}} d\mathbf{k} \sum_n f(E_{n\mathbf{k}} - \mu)\Omega_{n,yx}^{z}(\mathbf{k})$$

$$\Omega_{\mathrm{n,yx}}^{\mathrm{z}}(\mathbf{k}) = \mathrm{Im} \sum_{n' \neq n} \frac{\langle u_{n\mathbf{k}}|\hat{v}_y|u_{n'\mathbf{k}}\rangle\langle u_{n'\mathbf{k}}|\hat{v}_x|u_{n\mathbf{k}}\rangle - (x \leftrightarrow y)}{\left(E_{n\mathbf{k}} - E_{n'\mathbf{k}}\right)^2}$$

with a 201 × 201 × 201 $k$-points mesh.

**Acknowledgements**

The authors would like to thank Prof. Y.S. Wang in the Department of Physics and Astronomy at the University of Tennessee for instrument support. This work was supported by the Air Force Office of Scientific Research under grant no. FA9550-23-1-0502. The neutron part of this research used resources at the High Flux Isotope Reactor, a DOE Office of Science User Facility operated by the Oak Ridge National Laboratory. The beam time was allocated to the TAX (HB-3) instrument under proposal numbers IPTS-34782.1 and IPTS-35125.1.

**Data Availability Statement**

The data that support the findings of this study are available from the corresponding author upon reasonable request.

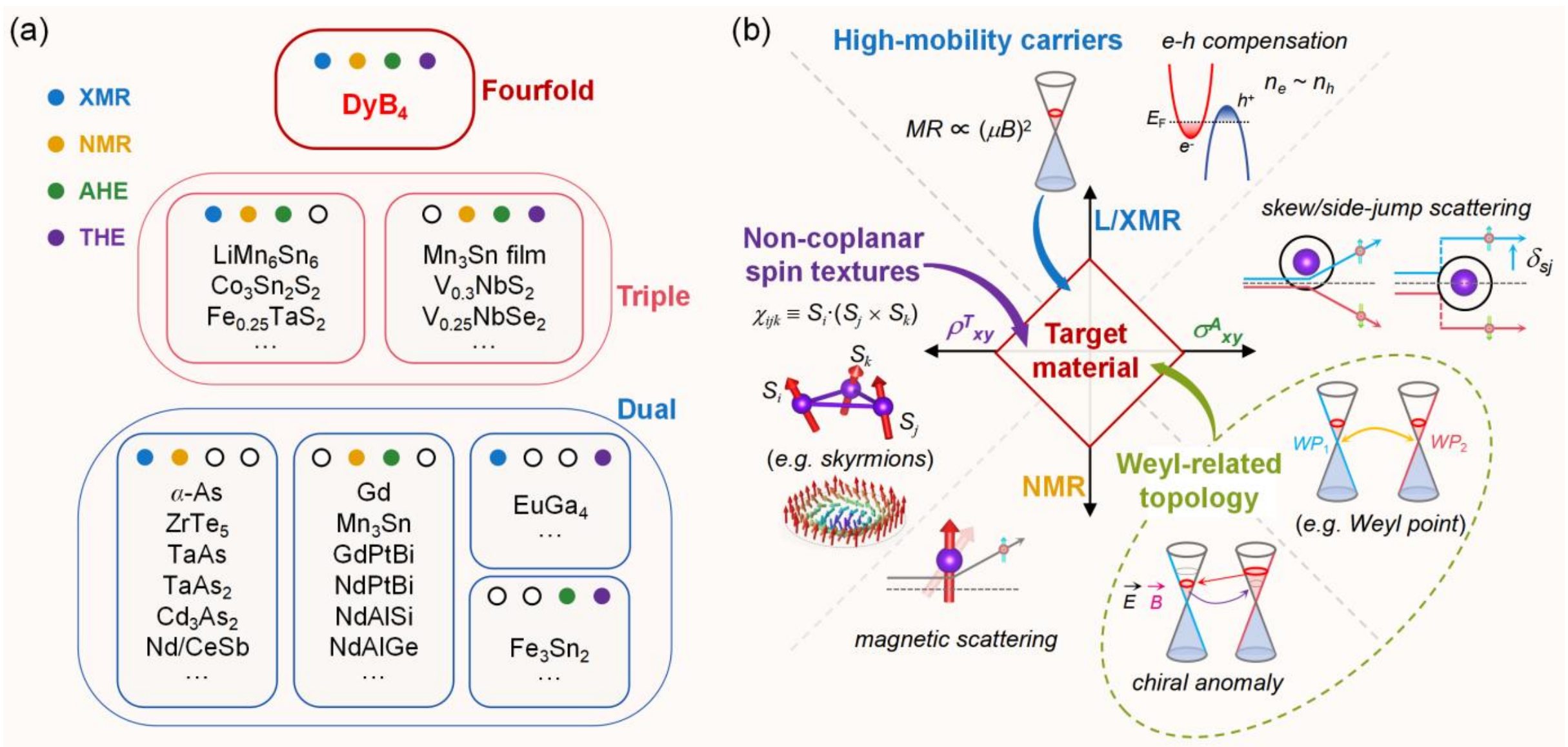

**Figure 1. Classification and design strategy of materials with coexisting anomalous magneto-transport responses.** (a) Classification of representative materials exhibiting coexisting XMR, NMR, AHE, and THE. Filled and open circles denote the presence and absence of each response, respectively. Only materials with two or more responses are shown. $DyB_4$, reported in this work, occupies the rare fourfold-combination regime. (b) Schematic illustration of the materials-design requirements for integrating large orbital magnetoresistance or extremely large magnetoresistance (L/XMR), chiral-anomaly-like NMR, AHE, and THE in a single material. High-mobility carriers or electron-hole compensation favor L/XMR, Weyl-related topology enables chiral-anomaly-like NMR and Berry-curvature-driven AHE, and non-coplanar spin textures generate topological Hall responses. Asymmetric scattering can further contribute to the AHE, whereas magnetic scattering may obscure XMR and intrinsic chiral-anomaly-like NMR. Thus, a promising target material for integrating all four anomalous magneto-transport responses is a magnetic Weyl material that combines high-mobility or electron-hole-compensated electronic states with tunable non-coplanar spin textures.

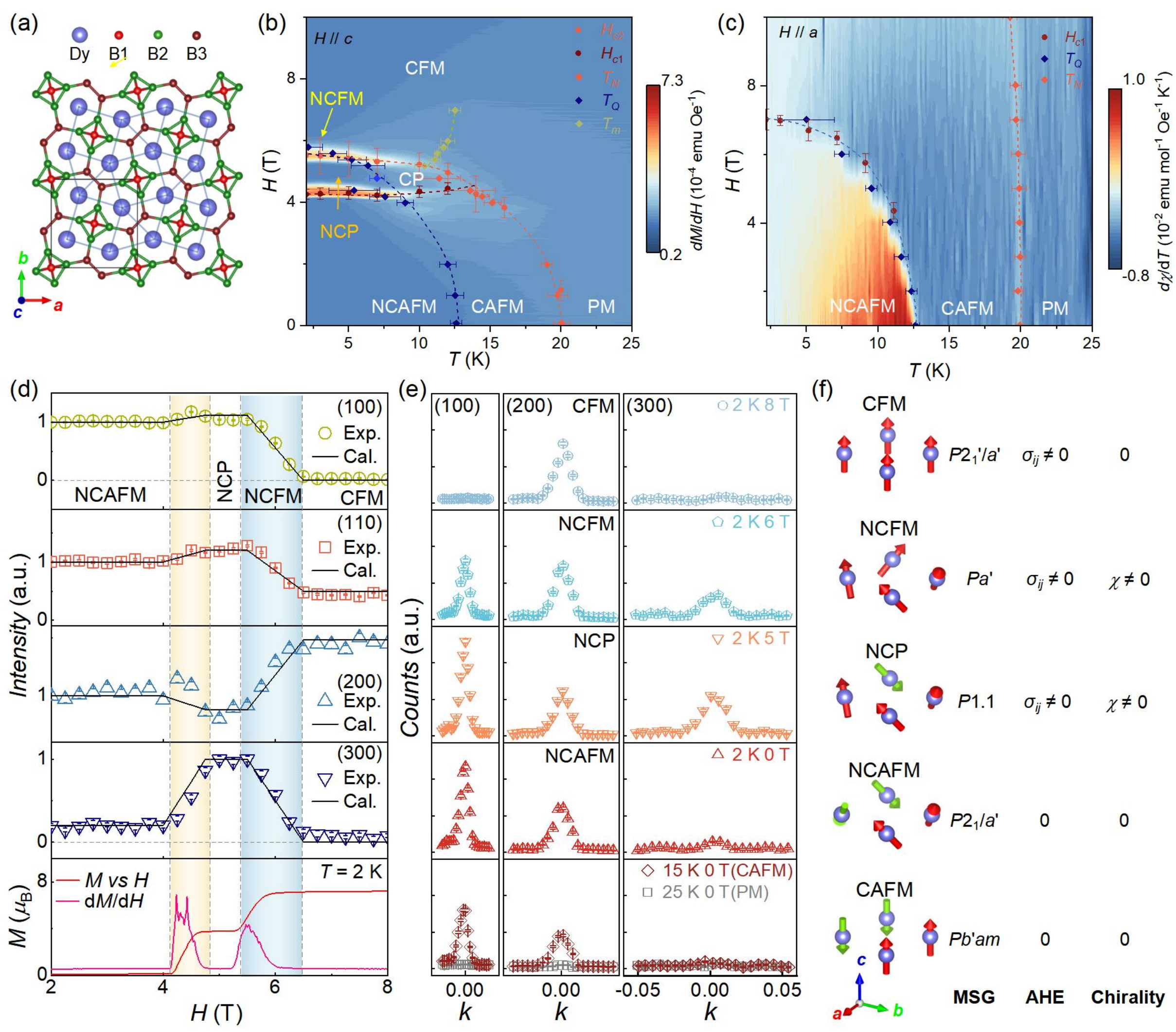

**Figure 2. Crystal structure and magnetic phase diagram with multiple magnetic states.** (a) Top view of the crystal structure of $DyB_4$, with Dy atoms on the SSL lattice. Magnetic field-temperature (*H*-*T*) phase diagrams under (b) out-of-plane (*H* // *c*) and (c) in-plane (*H* // *a*) magnetic fields. The symbols denote characteristic transition temperatures ($T_Q$, $T_N$, $T_M$) and critical magnetic field ($H_{c1}$, $H_{c2}$) extracted from the first derivative of the corresponding temperature-dependent magnetic susceptibility and field-dependent magnetization, respectively. (d) Observed *k*-scan of (*h*00) (*h* = 1, 2, 3) peaks in different magnetic states. (e) Magnetic-field dependence of the integrated intensities of the (100), (110), (200), and (300) magnetic diffraction peaks. All intensities are normalized by their integrated intensity at 2 K under 0 T, except for (300) normalized by its maximum. Field-dependent magnetization and its first-derivative curves are plotted for comparison. The shaded areas highlight the intermediate phases among the NCAFM, NCP, and CFM states. (f) Schematic illustration of the evolution of magnetic states in $DyB_4$ and the corresponding MSG, AHE tensor, and scalar spin chirality of each state.

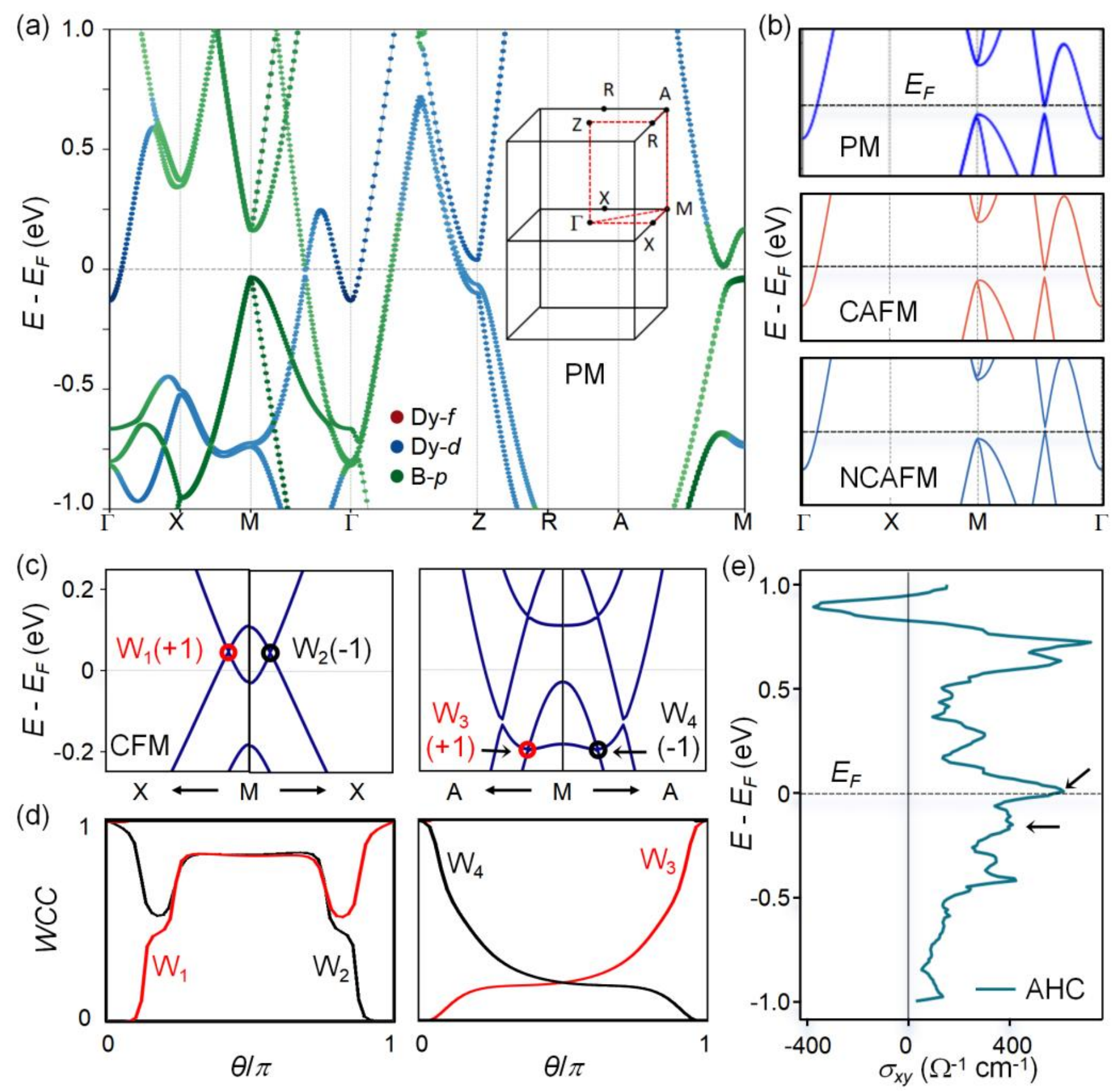

**Figure 3. Electronic structure of $DyB_4$ in different magnetic states and Weyl-related topology.** (a) Orbital-resolved band structure of $DyB_4$ in the PM state along the high-symmetry lines. The bands in different colors represent the components of Dy-4*d*, Dy-4*f*, and B-2*p* orbitals, respectively. The inset shows the bulk Brillouin zone of $DyB_4$. (b) Evolution of the band structure near $E_F$ along Γ-X-M-Γ while cooling down. (c) Expanded band structures near the selected band-crossing points along X-M-X and A-M-A paths in the CFM state, and (d) the corresponding evolution of the Wannier charge center (WCC) on the spheres centered at the selected Weyl points. The Weyl points of topological band crossings are represented by red and black dots, depending on their chiral charges. (e) Chemical-potential energy-dependent AHC ($\sigma^{A}_{xy}$) for $DyB_4$ in the CFM state. The arrows denote the energies of the corresponding Weyl points.

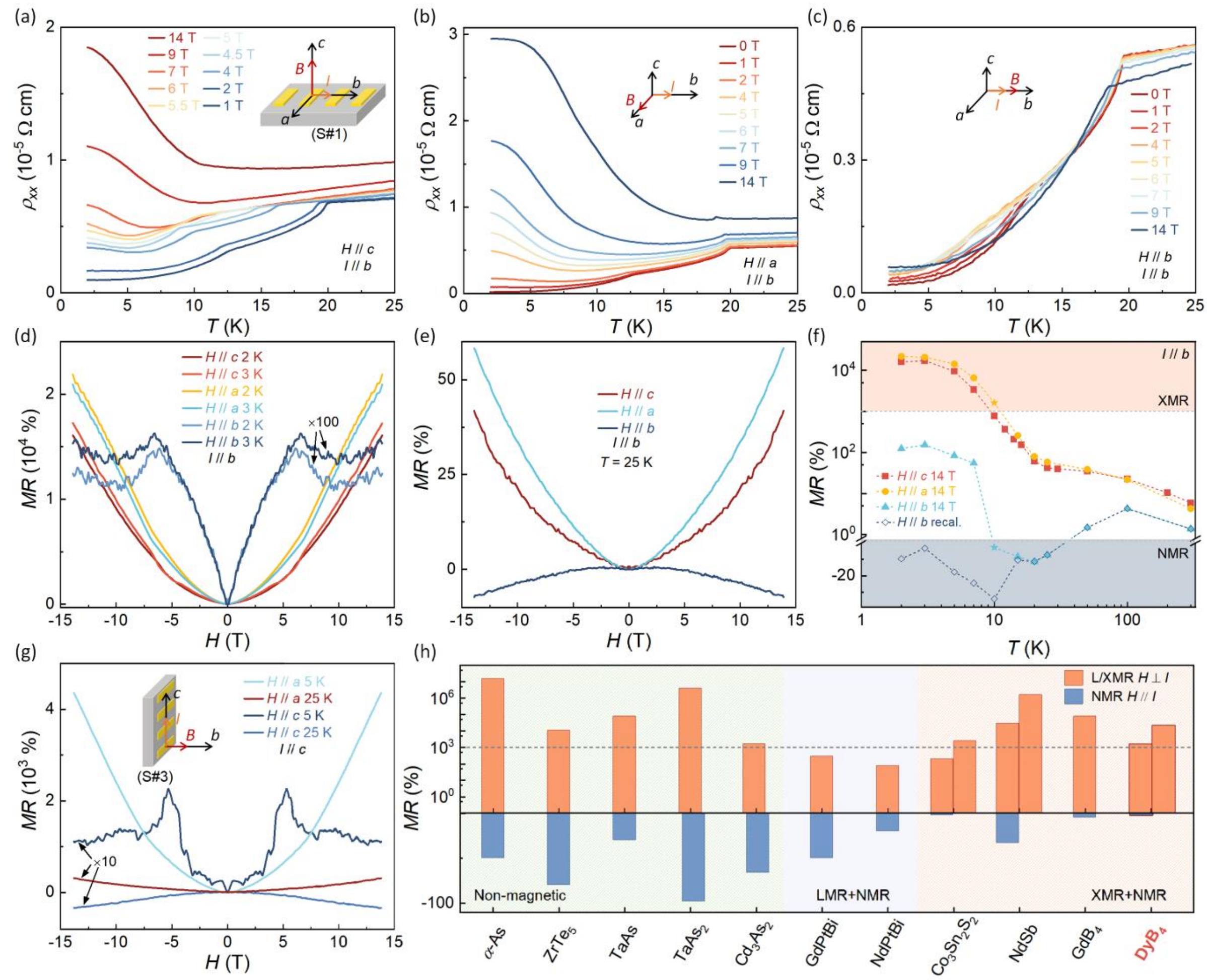

**Figure 4. Metal-insulator-like crossover, extremely large transverse MR, and chiral-anomaly-like NMR.** Temperature-dependent resistivity of $DyB_4$ under (a) out-of-plane magnetic fields (*H* // *c*), (b) in-plane magnetic fields perpendicular to the electrical current (*H* // *a*, *H* ⊥ *I*), (c) in-plane magnetic field parallel to the current (*H* // *b*, *H* // *I*). The inset illustrates the measurement configurations on a polished rectangular sample in the *ab*-plane (S#1), with the electrical current applied along the *b*-axis (*I* // *b*). MR curves measured under different magnetic field directions at (d) 2 K, 3 K, and (e) 25 K. (f) Temperature-dependent MR under a 14 T magnetic field applied along different directions. To highlight the NMR feature, the MR for the longitudinal configuration is recalculated by setting the maximum MR value at each temperature to zero (*H* // *b* recal.). The shaded regions indicate the NMR and XMR regimes, with the MR at 10 K (stars) showing the coexistence of both features. (g) XMR and NMR features at selected temperatures with the electrical current applied along the *c*-axis (*I* // *c*). The inset illustrates the measurement configurations on a polished rectangular sample in the *ac*-plane (S#3). (h) Comparison of representative materials showing the coexistence of L/XMR and chiral-anomaly-like NMR. The misalignment means that the L/XMR and NMR are observed at different temperatures.

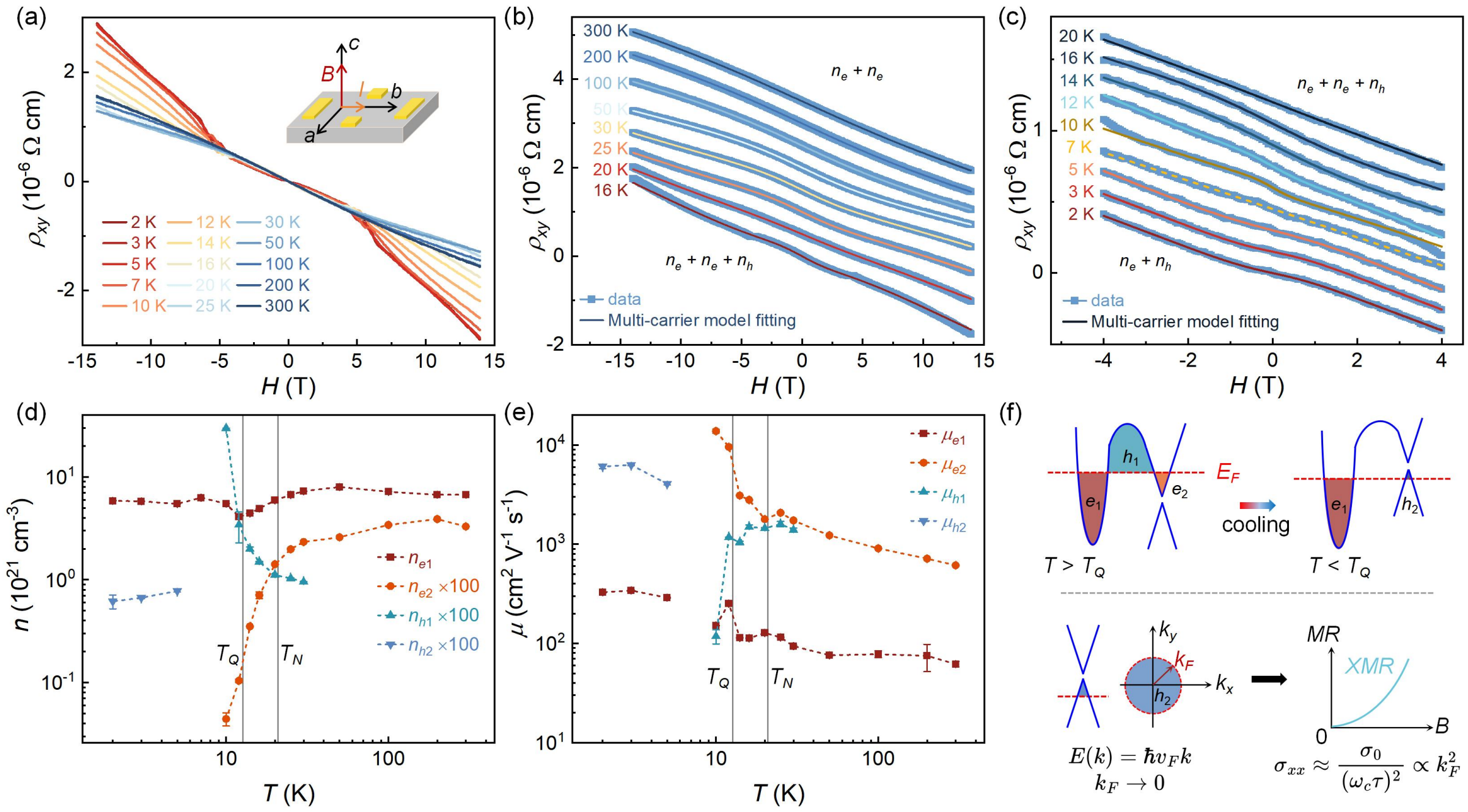

**Figure 5. Hall resistivity analysis based on a multi-carrier model.** (a) The field-dependent Hall resistivity at different temperatures. The inset shows the measurement geometry, where the electrical current and magnetic field follow the *b*- and *c*-axes, respectively. Analysis of the Hall resistivity in (b) the PM state, (c) the NCAFM and CAFM states under low magnetic field. Symbols represent the experimental data, and solid lines denote the fitting results based on the multi-carrier model. The data are vertically shifted for clarity. The temperature dependence of (d) the carrier density and (e) the mobility of each charge carrier. (f) Schematic of the band structure evolution when cooling across $T_Q$ and the possible mechanism of XMR, considering the steep linear dispersion and the emergence of the high-mobility small hole pocket.

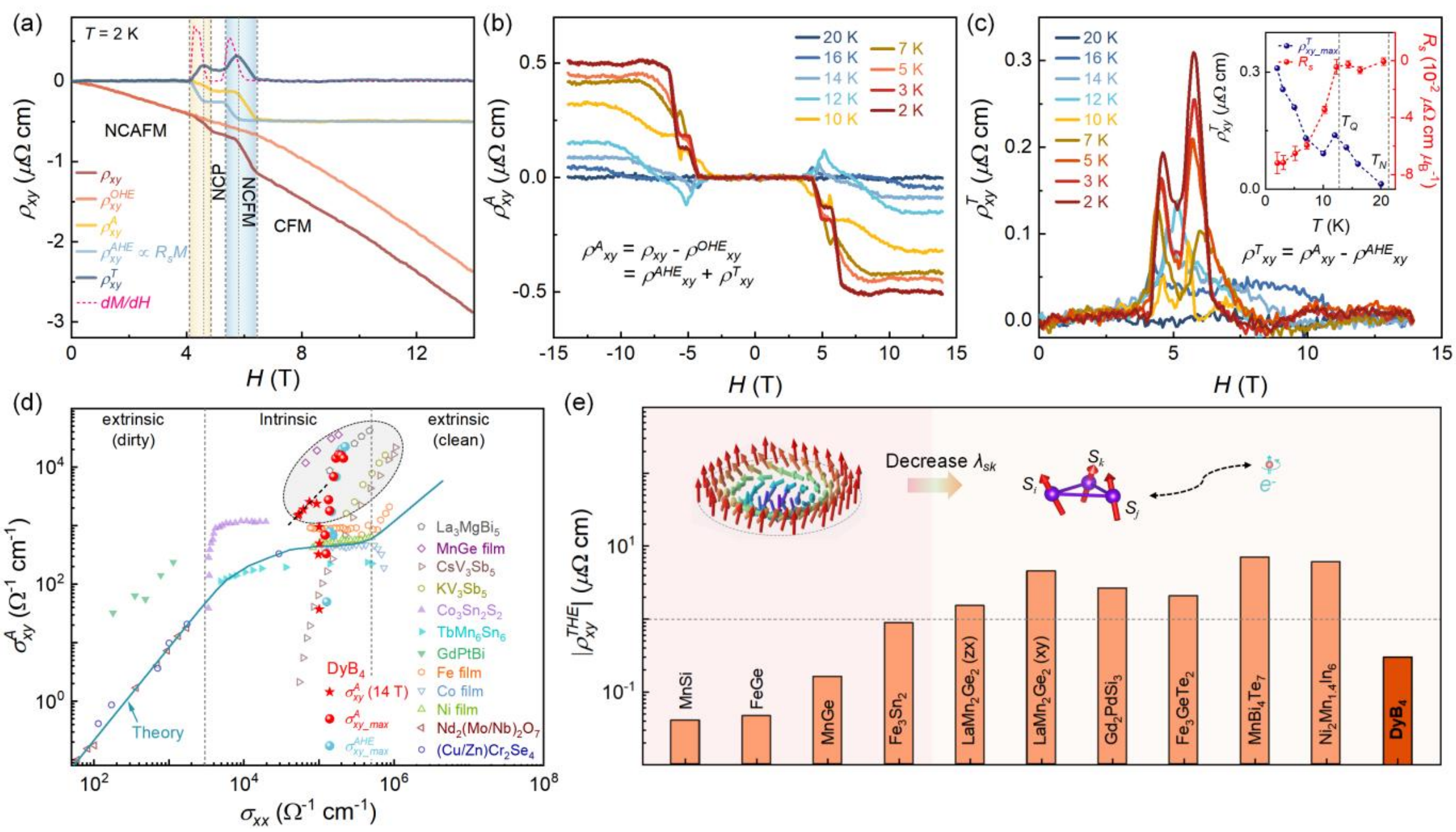

**Figure 6. Anomalous Hall resistivity and topological Hall resistivity.** (a) Decomposition of the ordinary, anomalous, and topological Hall resistivity at 2 K. The shaded areas show the intermediate phases among the NCAFM, NCP, and CFM states. (b) The combined anomalous and topological Hall term, $\rho^{A}_{xy}$, at different temperatures. (c) Magnetic-field dependence of the topological Hall resistivity, $\rho^{T}_{xy}$, at different temperatures. The inset shows the corresponding temperature dependencies of the anomalous Hall coefficient, $R_s$, and the maximum of the topological Hall resistivity, $\rho^{T}_{xy_max}$. (d) Comparison of the anomalous Hall response in various materials, shown as a logarithmic plot of AHC $\sigma_{xy}$ versus longitudinal conductivity $\sigma_{xx}$. The solid line represents the theoretical prediction. The plot is divided into three regimes: the poor-conducting regime ($\sigma_{xx} < 3 \times 10^3\ \Omega^{-1}\ \mathrm{cm}^{-1}$) dominated by extrinsic impurity scattering; the good-metal regime ($3 \times 10^3\ \Omega^{-1}\ \mathrm{cm}^{-1} < \sigma_{xx} < 5 \times 10^5\ \Omega^{-1}\ \mathrm{cm}^{-1}$) dominated by the intrinsic Berry curvature contribution; and the superclean region ($\sigma_{xx} > 5 \times 10^5\ \Omega^{-1}\ \mathrm{cm}^{-1}$) dominated by the extrinsic skew scattering. The shaded region highlights an unconventional AHC regime in which skew scattering cooperates with the intrinsic Berry curvature. The black dashed line shows the scaling relationship between AHC at 14 T, $\sigma^{A}_{xy}$ (14 T) (marked by stars), and the corresponding longitudinal conductivity. (e) Comparison of the maximum topological Hall resistivity $|\rho^{T}_{xy}|$ in $DyB_4$ with those of representative topological Hall materials. The inset shows the schematic of THE governed by scalar spin chirality and the coupling between conduction electrons and localized magnetic moments.